\documentclass[journal,twocolumn]{IEEEtran}
\usepackage{amsfonts}
\usepackage{times}
\usepackage{graphicx}
\usepackage{latexsym}
\usepackage{dsfont}
\usepackage{amssymb}
\usepackage{amsmath}
\usepackage{cite}
\usepackage{verbatim}
\usepackage{subfigure}

\newcommand{\figref}[1]{{Fig.}~\ref{#1}}

\def\bb0{{\mathbb{0}}}

\def\ba{{\mathbf{a}}}
\def\bb{{\mathbf{b}}}

\def\bg{{\mathbf{g}}}
\def\bh{{\mathbf{h}}}

\def\b0{{\mathbf{0}}}

\def\bbC{{\mathbb{C}}}

\def\sf0{{\mathsf{0}}}

\usepackage{epstopdf}
\usepackage{enumerate}
\usepackage{algorithmicx}
\usepackage{algorithm}
\usepackage{amsmath}
\usepackage[noend]{algpseudocode}
\usepackage{float}
\usepackage{hyperref}
\usepackage{color}
\usepackage{makeidx}
\usepackage{bbm}
\usepackage{graphicx}
\usepackage{balance}

\newcommand{\sref}[1]{{Section}~\ref{#1}}

\begin{document}

\title{Relay Aided Intelligent Reconfigurable Surfaces: Achieving the  Potential Without So Many Antennas}

\author{Xiaoyan Ying, Umut Demirhan, and Ahmed Alkhateeb \thanks{The authors are with the School of Electrical, Computer and Energy Engineering, Arizona State University, (Email: xying2, udemirhan, alkhateeb@asu.edu).}}
\maketitle

\begin{abstract}

This paper proposes a novel relay-aided intelligent reconfigurable surface (IRS) architecture for future wireless communication systems. The proposed architecture, which consists of two side-by-side intelligent surfaces connected via a \textit{full-duplex} relay, has the potential of achieving the promising gains of intelligent surfaces while requiring much smaller numbers of reflecting elements. Consequently, the proposed IRS architecture needs significantly less channel estimation and beam training overhead and provides higher robustness compared to classical IRS approaches. Further, thanks to dividing the IRS reflection process over two surfaces, the position and orientation of these surfaces can be optimized to extend the wireless communication coverage and enhance the system performance. In this paper, the achievable rates and required numbers of elements using the proposed relay-aided IRS architecture are first  analytically characterized and then evaluated using numerical simulations. The results show that the proposed architecture can achieve the data rate targets with much smaller numbers of elements compared to typical IRS solutions, which highlights a promising path towards the practical deployment of these intelligent surfaces. 
\end{abstract}

{\IEEEkeywords  Intelligent reconfigurable surfaces, full-duplex, relay, millimeter wave, MIMO.
}

\section{Introduction} \label{sec:Intro}

Intelligent reconfigurable surfaces (IRSs) are envisioned as intrinsic components of future wireless communication systems \cite{Hu2018,Taha2019,Basar2019,renzo2020smart}. This is mainly thanks to its promising gains in enhancing the coverage and  rates in future millimeter wave (mmWave) and terahertz networks where coverage is a major challenge. To achieve its potential gains, however, massive numbers of elements need to be deployed at these surfaces. This leads to critical challenges in terms of the channel estimation/beam training overhead as well as the robustness of these systems with very narrow beams, which may render the practical deployment of  these extremely large surfaces infeasible. This paper proposes a novel relay-aided IRS architecture that has the potential of achieving the IRS promising gains with much less numbers of elements; opening the door for realizing these surfaces in practice. 

Multi-antenna technologies such as mmWave  Multiple-Input-Multiple-Output (MIMO) \cite{Alkhateeb2014,HeathJr2016,Roh2014,Alkhateeb2018,Li2020a} and massive MIMO {\cite{Larsson2014a,Lu2014,Marzetta2010,Zhang2020a,Alrabeiah2019}} are key technologies for meeting the data rates demands in 5G and future wireless systems. As these systems move to higher frequency bands, though, they become more susceptible to blockages. This poses a critical challenge for future wireless communication networks. The concept of employing intelligent reconfigurable surfaces  to overcome these blockages and provide alternative high-quality paths between the transmitters and receivers has been recently proposed and attracted significant interests \cite{Hu2018,Huang2019,Taha2019,taha2020deep}. These architectures consist of massive numbers of nearly-passive elements, which are configured to reflect/focus the power towards the intended receivers. 
While these surfaces have many promising applications, very large numbers of  reflect elements are needed to achieve sufficient receive power, leading to high training overhead and extremely narrow beams, which makes the precise control of steering direction critical. 

A more widely accepted method for adapting wireless communication environment is by using relay stations, which may also generate additional wireless routes toward the destination. While both relays and intelligent surfaces are relatively similar, the relay plays the role of receiving and retransmitting the signal with amplification. In \cite{renzo2019reconfigurable,Bjoernson2019a}, comparisons were made between intelligent surfaces and decode-and-forward(DF)/amplify-and-forward(AF) relays, reaching the conclusion that an IRS needs hundreds of reconfigurable elements to be competitive against  relays. However, conventional relays are lacking the ability of focusing the signal, which limits its ability for wireless coverage and increases the interference to unintended receivers. Further,  MIMO relays are costly and bulky with high power consumption.

In this paper, we propose a novel IRS architecture that consists of two IRS surfaces connected via a full-duplex relay. The proposed architecture merges the gains of both relays and reconfigurable surfaces and splits the required SNR gain between them. This  architecture can then significantly reduce the required number of elements while achieving the same spectral efficiencies.  Consequently,  the proposed architecture needs much less channel estimation/beam training overhead and provides enhanced robustness compared to traditional IRS solutions. An important aspect of the proposed architecture is splitting the reflection process over two intelligent surfaces. This allows leveraging full-duplex relays with practical isolation. Further, this enables the proposed architecture to be deployed in very flexible ways by optimizing the position and orientation of the two surfaces, which leads to much better coverage. After describing the proposed architecture, this paper develops an accurate mixed near-far field channel model that describes the composite channel between the transmitter/receiver and the relay through the IRS surfaces. Further, the paper derives closed-form expressions for the achievable rates using the proposed relay-aided IRS architecture with AF and DF relays.
 Finally, these rates are evaluated using numerical simulations which further highlight the promising gains of the proposed architecture. 



\begin{figure}[t]
	\centering
	\includegraphics[width=1\columnwidth]{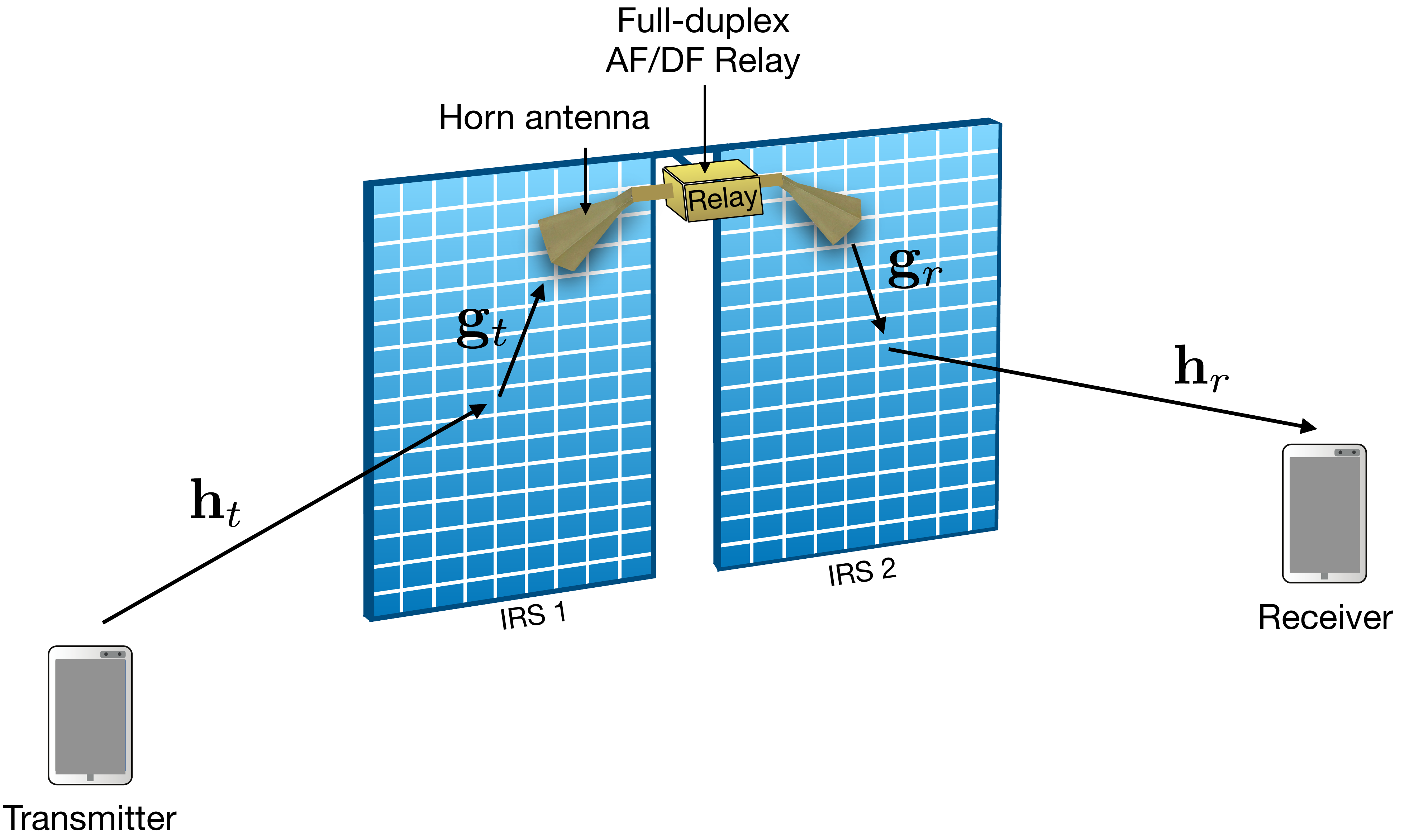}
	\caption{The proposed relay-aided IRS architecture consists of two intelligent reconfigurable  surfaces connected via a full-duplex relay. IRS 1 reflects the transmitter's signal towards the relay and IRS 2 reflects signal transmitted by the relay towards the target receiver. }
	\label{fig:sysdiag}
\end{figure}

\section{Proposed Architecture: Relay-Aided IRS}

Intelligent reconfigurable surfaces have the potential of enhancing the coverage and data rates of future wireless communication systems. This is particularly important for mmWave and Terahertz systems where network coverage is a critical problem. The current approach in realizing these surfaces is through using massive numbers of nearly-passive elements that \textit{focus} the incident signals towards the desirable direction. In order to achieve sufficient receive power, however, these surfaces will typically need to deploy tens of thousands of antenna elements (as will be shown in \sref{sec:Results}). \textbf{Having intelligent reconfigurable surfaces with that many antennas carries fundamental problems that may render these surfaces infeasible.} In addition to the high-cost in building them, these surfaces have extremely narrow beams which incur massive training overhead with which supporting even low-mobility applications is questioned. Further, narrow beams constitute a critical challenge for the robustness of the communication links as any little movement may result in a sudden drop in the receive power. With the motivation of overcoming these challenges and enabling the potential gains of intelligent reconfigurable surfaces in practice, we propose a novel architecture based on merging these surfaces with full-duplex relays. Next, we  briefly describe the proposed architecture and highlight its potential gains. 

\subsection{Architecture Description}

The core idea of the proposed architecture is to make the intelligent surfaces capable of amplifying the power of the incident signals without the need to explicitly deploy power amplifiers at the elements of these surfaces. This has the potential of splitting the required SNR gain between the array gain (using the focusing capability of the IRS) and the power amplification gain.  \textbf{To achieve this goal, we propose the simple architecture shown in \figref{fig:sysdiag}, where two intelligent reconfigurable surfaces are connected via a full-duplex relay.} This architecture operates as follows: When the transmitter transmits a signal, the first intelligent surfaces (IRS 1) reflects this signal towards the horn antenna of the attached relay. This \textit{full-duplex} relay then amplifies (or decodes) the signal and retransmits it towards the second intelligent surface (IRS 2). Finally, this surface reflects and focuses the signal towards the target receiver. The two surfaces switch their roles as the direction of communication switches.  Note that a key aspect of the proposed architecture is having two surfaces doing different (transmit/receive) functions at any point in time. This allows employing a full-duplex relay (with reasonable isolation) and enables the proposed relay-aided IRS architecture to continuously reflect the incident signals. 

\begin{figure}[t]
	\centering
	\includegraphics[width=1\columnwidth]{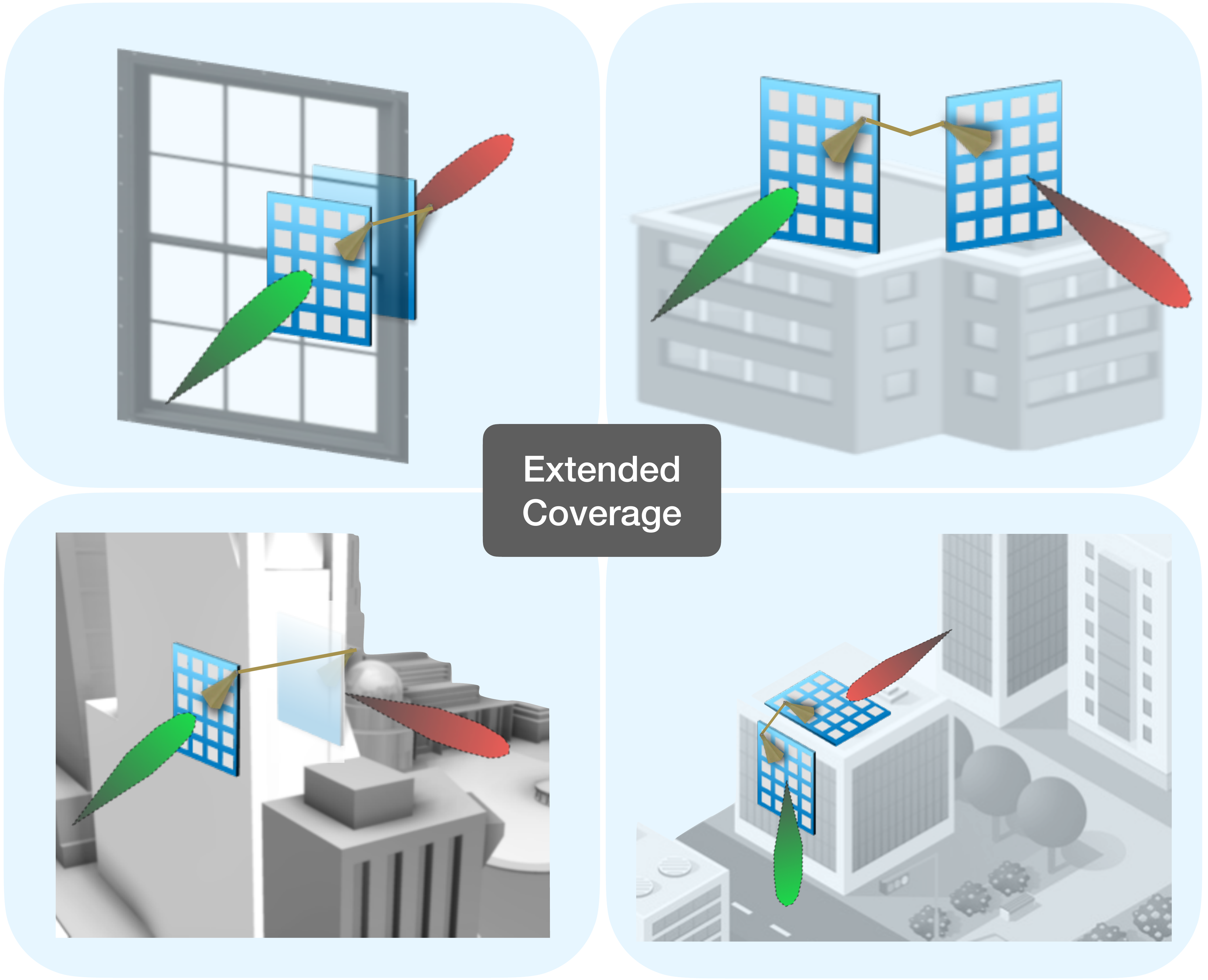}
	\caption{The proposed relay-aided IRS architecture has the potential of extending the wireless communication coverage by configuring the position and orientation of the two intelligent surfaces.}
	\label{fig:coverage}
\end{figure}

\subsection{Motivation and Potential Gains}
The proposed relay-aided IRS architecture has several potential gains compared to the classical IRS architecture that has a single surface. Next, we briefly highlight these gains. 
\begin{itemize}
	\item  \textbf{Less number of elements:} To achieve a sufficient SNR gain, the proposed architecture has the possibility to split this required gain between the relay power amplification gain and the focusing gain of the intelligent surface. This can considerably reduce the required number of elements at the intelligent surfaces.

	\item \textbf{Low beam training overhead:} To realize the potential beamforming gain, the controllable elements  of the intelligent surfaces need to be configured based on the channels between these surfaces and the transmitters/receivers. Acquiring this channel knowledge (or equivalently finding the best beam), however, requires huge training overhead in classical intelligent reconfigurable surfaces that employ massive numbers of elements. This imposes a critical challenge for the feasibility of these surfaces in practical deployments. Given that the proposed architecture has the potential of achieving the same SNR gains with much less number of elements (and hence much less training overhead), it presents an interesting path for realizing these systems in practice. 

	\item \textbf{Wider beams for higher robustness:} Another critical challenges that follow from employing a massive number of elements in classical reconfigurable surfaces is the very small beamwidth of the focusing beams. These laser-like beams highly affect the robustness of these systems as the links can be abruptly disconnected with any small movement by the transmitter or the receiver. In contract, and thanks to requiring a smaller number of elements, the proposed relay-aided IRS architecture employ wider beams, which enhances the robustness of the system. 

	\item \textbf{Better coverage:} An interesting characteristic of the proposed relay-aided IRS architecture is the use of two intelligent surfaces. This allows moving those two surfaces to extend the communication coverage and overcome potential blockages. In \figref{fig:coverage}, we demonstrate some candidate deployment scenarios that highlight the potential of the proposed relay-aided IRS architecture in extending the coverage in wireless networks. 
	
\end{itemize}

\section{System Model} \label{sec:system}

Consider the communication system shown in \figref{fig:sysdiag} where a transmitter and receiver are communicating through the proposed relay-aided intelligent reconfigurable surface. For simplicity, we assume that there is no direct line-of-sight link between the transmitter and receiver (assuming this link is either blocked or negligible). Further, we adopt a scenario where the transmitter and receiver have single antennas. The proposed model and results in this paper, however, can be extended to the case with multi-antenna transceivers. When the transmitter sends the signal $s$, this signal is first reflected by the receive reflecting surface (IRS 1 in \figref{fig:sysdiag}) to the relay receiver antenna. This signal is then amplified (in the case of AF relay) or regenerated (in the case of DF relay) before being transmitted to the second reflecting surface (IRS 2 in \figref{fig:sysdiag}), which reflects the signal towards the receiver. 
\textbf{It is important to note here that since the two reflect arrays are separated for receiving and transmitting purposes, the proposed relay-aided IRS architecture can efficiently operate in a full-duplex mode}, with reasonable isolation between the  directional transmit and receive antennas of the relay. This allows the proposed IRS architecture to work on continuously reflecting the incident signals without requiring additional time resources. 

Assume that each intelligent surface has $M$ antennas, and let $\bh_{t}  \in \mathbb{C}^{M \times 1}$ and $\bg_{t}  \in \mathbb{C}^{M \times 1}$ denote the channels between the transmitter and IRS 1, and between IRS 1 and the relay receive antenna. Then, the receive signal at the relay can be written as 
\begin{equation} \label{eq:receive1}
y_\mathrm{RIR}= \sqrt{p_\mathrm{t}} \ \bg_t^T \mathbf{\Psi}_1 \bh_t s + n_1,
\end{equation}
where $p_\mathrm{t}$ denotes the transmit power at the transmitter, $s$ is the transmit symbol with unit average power, and $n_1 \sim \mathcal{N} \left( 0, \sigma_{1}^2 \right)$ is the receive noise at the relay. The $M \times M$ diagonal matrix $ \mathbf{\Psi}_1$ is the interaction matrix of the first intelligent reconfigurable surface (IRS 1). If $\boldsymbol{\psi}_1$ denotes the diagonal vector of  $\boldsymbol{\Psi}_1$, i.e., $\boldsymbol{\Psi}_1=\mathrm{diag}\left({\boldsymbol{\psi}_1}\right)$, then we can rewrite \eqref{eq:receive1} as 
\begin{equation} \label{eq:receive2}
y_\mathrm{RIR}= \sqrt{p_\mathrm{t}} \ \left(\bh_t \odot \bg_t \right)^T \boldsymbol{\psi}_1 s + n_1,
\end{equation}
where $\odot$ is the Hadamard product. In this paper, we  focus on the case when the intelligent surfaces interact with the incident signals via phase shifters, i.e., $\boldsymbol{\psi}_1=\sqrt{\kappa_1} \left[e^{j \phi^1_1}, ..., e^{j \phi^1_M}\right]$, with $\kappa_1$ representing the power reflection efficiency of the first intelligent surface.  At the relay, the receive signal is processed by either applying an amplification gain (for the case of AF relay) or decoding followed by retransmission (for DF relays). 

\textbf{For AF relays:} An amplification gain $\beta$ will be applied to the receive signals before retransmitting it towards the second intelligent reconfigurable surface (IRS 2). This surface will then reflect the signal to the receiver using its interaction matrix $\boldsymbol{\Psi}_2$, defined similarly to $\boldsymbol{\Psi}_1$. If $\bg_{r}  \in \mathbb{C}^{M \times 1}$ and $\bb_{r}  \in \mathbb{C}^{M \times 1}$ represent the channels between IRS 2 and the relay transmit antennas and between IRS 2 and the receiver, then the receive signal at the receiver can then be written as 
\begin{equation}
y_{r}= \sqrt{\beta }  \left(\bh_r \odot \bg_r\right)^T \boldsymbol{\psi}_2 \left(\sqrt{p_t} \left(\bh_t \odot \bg_t \right)^T \boldsymbol{\psi}_1 s  + n_1 \right)+ n_2,
\end{equation}
where $n_2  \sim \mathcal{N} \left( 0, \sigma_{2}^2 \right)$ is the receive noise at the receiver.

\textbf{For DF relays:} The receive signals will be decoded and retransmitted with power $p_r$ to the second intelligent surfaces  (IRS 2), which reflects the signal towards the receiver using its interaction matrix $\boldsymbol{\Psi}_2$. In this case, the receive signal at the receiver can be written as 
\begin{equation}
y_r=\sqrt{p_r}  \left(\bh_r \odot \bg_r\right)^T \boldsymbol{\psi}_2  s + n_2.
\end{equation}

\noindent An important note on the transmit and receive side composite channels,  $ \left(\bh_t \odot \bg_t \right)$ and $ \left(\bh_r \odot \bg_r\right)$, is that they combine far-field channels $\bh_t, \bh_r$ and near-field channels $\bg_t, \bg_r$. In the next section, we develop an accurate model for these channels.

\begin{figure}[t]
	\centering
	\includegraphics[width=.9\columnwidth]{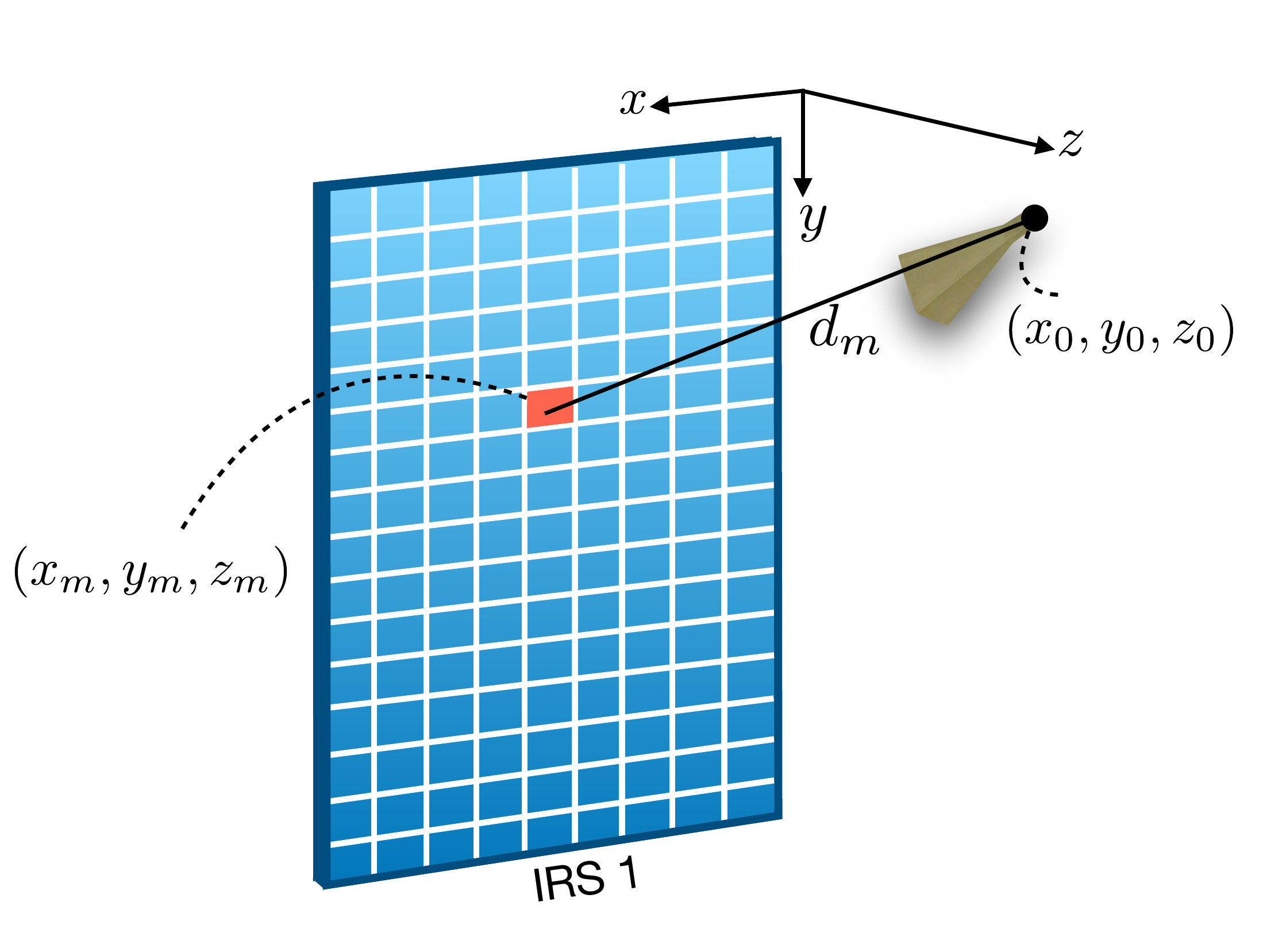}
	\caption{The channels between the intelligent surfaces and the relay, $\bg_t$ and $\bg_r$, adopt near-field and spherical wave modeling. }
	\label{fig:channel}
\end{figure}

\section{Channel Model: Mixed Near-Far Field Model}

One important characteristic of the proposed relay-aided IRS architecture is that the channels between intelligent surfaces and the transmitter/receiver can be modeled as far-field channels while the channels between the surfaces and the relay need to adopt near-field modeling. In this section, we will describe in detail the composite channel model for the transmit side, which we denote $\bh^\circ_t=\bh_t \odot \bg_t$. The  receive-side composite channel $\bh^\circ_r=\bh_r \odot \bg_r$ can be similarly defined.

Given the description of the relay-aided IRS architecture in \sref{sec:system}, we can write the transmit-side composite channel as 
\begin{equation}
	\bh^\circ_t=\bh_t \odot \boldsymbol{\varsigma}_t \odot \boldsymbol{\Theta}_t,
\end{equation}
where $\boldsymbol{\varsigma}_t$ and $\boldsymbol{\Theta}_t$ are the magnitude and phase vectors of the near-field IRS-relay channel $\bg_t$, i.e., $\bg_t=\boldsymbol{\varsigma}_t \odot \boldsymbol{\Theta}_t$.  First, we describe the far-field channel vectors, $\bh_t$, using a geometric channel model \cite{Taha2019}. In this model, the signal propagating between the transmitter and IRS 1 experiences $L$ clusters, and each cluster contributes with one ray via a complex coefficient $\alpha_{\ell} \in \bbC$ and azimuth/elevation angles of arrival, $\theta^{\mathrm{az}}_{\ell},\theta^\mathrm{el}_{\ell} \in [0, 2 \pi)$. Hence, the channel $\bh_t$ can be written by 
\begin{equation} \label{eq:channelmodel}
\bh_t=\sum_{\ell=1}^L  \sqrt{\rho_t}  \alpha_\ell  \ba\left(\theta^{\mathrm{az}}_{\ell},\theta^\mathrm{el}_{\ell}\right) 
\end{equation}
where $\rho_t$ denotes the path loss between the transmitter and the IRS 1, and  $\ba(.) \in \mathbb{C}^{M \times 1}$ represents the  array response vector of the first intelligent surface (IRS 1).

For the channel between the intelligent reconfigurable surface and the horn antenna, given the small distance between them, near-field and spherical propagation models need to be considered \cite{Bohagen2007,Cheng2018}. Near-field effects are reflected on both the magnitude and phase of the channel entries and magnitude depends on the free-space path-loss, the polarization mismatch and the effective aperture area of the antenna. For IRS elements of side-length $\frac{\lambda}{2}=\frac{f}{2c}$, the magnitude of the channel between element $m$ of IRS 1 and the relay antenna, $\left[\boldsymbol{\varsigma}_t\right]_m$, can be approximated as \cite{bjrnson2020power}
\begin{equation}
\left[\boldsymbol{\varsigma}_{t}\right]_{m}= \left(\frac{G_t}{4 \pi} \sum_{x \in \mathcal{X}} \sum_{y \in \mathcal{Y}} \left( \frac{C_{x,y}}{3 \left(\frac{y^2}{d^2}+1\right)}   + \frac{2}{3} \tan^{-1} C_{x,y} \right)\right)^{\frac{1}{2}}, 
\end{equation}
with 
\begin{align}
C_{x,y}=\frac{\frac{x y}{d^2}}{\sqrt{\frac{x^2}{d^2}+\frac{y^2}{d^2}+1}},
\end{align}
where $\mathcal{X}=\left\{\frac{c}{4 f \sqrt{\pi}}+ x_m - x_0, \frac{c}{4 f \sqrt{\pi}}- x_m + x_0\right\}$ and $\mathcal{Y}=\left\{\frac{c}{4 f \sqrt{\pi}}+ y_m - y_0, \frac{c}{4 f \sqrt{\pi}}- y_m + y_0\right\}$, with $c$ and $f$ denoting the speed of light and carrier frequency. The height of the relay antenna is denoted by $d=|z_m-z_0|$ and the gain of the horn antenna over the isotropic antenna is represented by $G_t$. Finally, following the spherical wave equations  \cite{Bohagen2007,Cheng2018}, the phase factor of the channel between the $m$th element of IRS 1 and the relay antenna, which is captured in the $m$th element of $\boldsymbol{\Theta}_t$, can be written as
\begin{equation}
\left[\boldsymbol{\Theta}_t\right]_m= e^{j \frac{2 \pi}{\lambda} \sqrt{(x_m-x_0)^2+(y_m-y_0)^2+d^2}},
\end{equation}
where $\lambda$ is the wavelength.

\section{Achievable Rates} \label{sec:rates}
In this section, we investigate the achievable spectral efficiency using the proposed relay-aided intelligent reconfigurable surfaces. First, we briefly review the spectral efficiency achieved by the standard intelligent reconfigurable surfaces and AF/DF relays. Then,  we derive the spectral efficiency of the proposed relay-aided IRS architecture with both AF and DF relays, respectively. In the following derivations, we assume that perfect channel state information is available at IRS, relay, and relay aided IRS. Extending these results to account for imperfect channel state information can be an interesting direction for future extensions.

\subsection{intelligent reconfigurable Surfaces}

We start by deriving spectral efficiency of IRS for comparison purposes. By adopting the same channel definitions for the transmitter-IRS and IRS-receiver channels, i.e., $\bh_t$ and $\bh_r$, we formulate the received signal as
\begin{equation}
y_r =  \sqrt{p_\mathrm{t}} \ \bh_r^T \mathbf{\Psi} \bh_t s + n_2, \label{eqn:IRSsignalmodel}
\end{equation}
where $n_2$ is the receiver noise as defined previously, and $\mathbf{\Psi}=\mathrm{diag}(\boldsymbol{\psi})$ is the interaction matrix of the IRS with $\boldsymbol{\psi}=\sqrt{\kappa} \left[e^{j \phi_1}, ..., e^{j \phi_M}\right]$. For the spectral efficiency of IRS, we can write
\begin{align}
R_{\mathrm{IRS}} &= \ \max_{\boldsymbol{\psi}} \ \log_2\bigg(1+\frac{p_t \kappa |\left(\bh_t \odot \bh_r \right)^T \boldsymbol{\psi}|^2} {\sigma_2^2} \bigg) \label{eqn:irsopt1} \\ 
&= \log_2\bigg(1+\frac{p_t \kappa (\sum_{m=1}^M |[\bh_{t}]_m| |[\bh_{r}]_m|)^2}{\sigma_2^2} \bigg) \label{eqn:irsopt2}\\
&= \log_2\bigg(1+\frac{p_t \kappa M^2 \xi_{t,r}} {\sigma_2^2} \label{eqn:irsopt} \bigg) \\
&\leq \log_2\bigg(1+\frac{p_t \kappa M^2 \zeta_t \zeta_r} {\sigma_2^2} \bigg) \label{eqn:irsupperbound}.
\end{align}
Note that \eqref{eqn:irsopt1} is obtained by a transformation of \eqref{eqn:IRSsignalmodel} similar to \eqref{eq:receive1} and \eqref{eq:receive2}. In \eqref{eqn:irsopt2}, the IRS is configured to maximize the gain via applying inverse phase shift of combined receive and transmit channels such that $\phi_m^3=-\angle{[\bh_t]_m [\bh_r]_m}$. The results in \eqref{eqn:irsopt} are in a compact form by defining $\xi_{t,r}=(\frac{1}{M} \sum |[\bh_t]_m| |[\bh_r]_m|)^2$. Moreover, it can be upper-bounded with Cauchy-Schwarz inequality as given in \eqref{eqn:irsupperbound} with the definitions $\zeta_t=\frac{1}{M} \sum |[\bh_t]_m|^2$ and $\zeta_r=\frac{1}{M} \sum |[\bh_r]_m|^2$.

\textbf{LOS scenario:} The expression in \eqref{eqn:irsopt} can be further simplified in the case where only LOS path is available. In this case, channel between the transmitter and the IRS follows \eqref{eq:channelmodel} for $L=1$ and $\alpha_1=1$ resulting in $\bh_t=\sqrt{\rho_t} \ba\left(\theta^{\mathrm{az}}_{\ell},\theta^\mathrm{el}_{\ell}\right)$. Hence, $\xi_{t,r}=\rho_t \rho_r$ and
\begin{equation} \label{eqn:LIS}
R_\mathrm{IRS} = \log_2 \left(1+ { {p_t  \kappa M^2 \rho_t \rho_r }\over \sigma_2^2} \right)
\end{equation}
which is a similar expression to the upper-bound defined in \eqref{eqn:irsupperbound}, however, the equality is exactly satisfied with the scalar channel gain values $\rho_t$ and $\rho_r$.

\subsection{Standard Relays}
We also consider a standard relay with a single antenna in each direction, again adopting the same channel definitions $h_t$, $h_r$ for $M=1$. The spectral efficiency of the relay models follows the derivations of the classical work \cite{Laneman2004} with trivial changes due to (i) the absence of LOS channel between the transmitter and source, and (ii) the full-duplex operation without any interference. 

\subsubsection{DF Relay} With the given definitions, spectral efficiency of the DF relay can be written by
\begin{equation} \label{eqn:df-relay}
R^\mathrm{DF}_\mathrm{Relay} = \log_2 \left(1+ \min\left\{  \frac{p_t \zeta_t}{\sigma_1^2}, {p_r \zeta_r \over \sigma_2^2} \right\} \right),
\end{equation}
which simply selects the minimum rate of two channels utilized in the transmission.

\subsubsection{AF Relay} For AF operation, the relay amplifies the received signal with the amplifying coefficient $\beta$, leading to
\begin{equation} \label{eqn:af-relay}
R^\mathrm{AF}_\mathrm{Relay} = \log_2 \left(1+ \frac{p_t \beta \zeta_t \zeta_r }{\beta \zeta_r \sigma_1^2 + \sigma_2^2} \right).
\end{equation}
Note that the relay is subject to a power constraint $p_r$, resulting in constraint $\beta \leq \frac{p_r}{p_t \zeta_t + \sigma_1^2}$. For the equality where full power is applied by the relay, the expression can be further simplified to
\begin{equation} \label{eqn:af-relay2}
\bar{R}^\mathrm{AF}_\mathrm{Relay} = \log_2 \left(1+ \frac{\frac{p_t \zeta_t}{\sigma_1^2} \cdot \frac{p_r \zeta_r}{\sigma_2^2} }{\big(\frac{p_t \zeta_t}{\sigma_1^2} + \frac{p_r \zeta_r}{\sigma_2^2} + 1 \big)} \right).
\end{equation}

\subsection{Relay-Aided IRS}

Recall that relay-aided IRS can adopt either DF or AF operations depending on application. For instance, DF relay is preferable for frequency selective fading channels, while AF relay is favored when less transmission latency between base station and user is required. 

We take IRS gains equal as they are identical, i.e., $\kappa_1=\kappa_2=\kappa$.
To derive spectral efficiency of relay-aided IRS, we start with transmitter-relay direction, and write
\begin{align}
R_t &= \ \max_{\boldsymbol{\psi}_1} \ \log_2\bigg(1+\frac{p_t \kappa |\left(\bh_t \odot \bg_t \right)^T \boldsymbol{\psi}_1|^2} {\sigma_1^2} \bigg) \label{eq:rirtrcap} \\ 
&= \ \max_{\boldsymbol{\psi}_1} \ \log_2\bigg(1+\frac{p_t \kappa |\left(\bh^\circ_t \right)^T \boldsymbol{\psi}_1|^2} {\sigma_1^2} \bigg) \\ 
&= \max_{\phi^1_1,\ldots,\phi^1_M} \log_2\bigg(1+\frac{p_t \kappa (\sum_{m=1}^M [\bh^\circ_{t}]_m \, e^{j \phi^1_m} )^2} {\sigma_1^2} \bigg) \\ 
&= \log_2\bigg(1+\frac{p_t \kappa M^2 \xi_{t}^\circ}{\sigma_1^2} \bigg) \label{eqn:opt_1}
\end{align}
where \eqref{eqn:opt_1} is obtained by setting $\phi_m^1=-\angle{[\bh^\circ_{t}]_m}$ maximizing the expression, and defining $\xi_{t}^\circ = (\frac{1}{M} \sum |[\bh^\circ_{t}]_m|)^2$.

By applying the same operations in \eqref{eq:rirtrcap}-\eqref{eqn:opt_1}, we can write the spectral efficiency of relay-receiver direction by
\begin{align}
R_r&= \log_2\bigg(1+\frac{p_t \kappa M^2 \xi^\circ_r} {\sigma_2^2} \bigg)
\end{align}
with the phase shift values of IRS 2 being selected as $\phi_m^2=\angle{[\bh^\circ_{r}]_m}$ and $\xi_{r}^\circ = (\frac{1}{M} \sum|[\bh^\circ_{r}]_m|)^2$.

\subsubsection{DF Relay Operation} In a similar way to \eqref{eqn:df-relay}, DF-relay-assisted IRS supported wireless network can support the spectral efficiency
\begin{align} \label{eqn:RIR-DF}
\begin{split}
R_\mathrm{RIR}^\mathrm{DF} & = \min\{R_t, R_f\} \\
& = \log_2\bigg(1+ \kappa M^2 \cdot \min\left\{\frac{p_t \xi^\circ_t} {\sigma_1^2}, \frac{p_r \xi^\circ_r} {\sigma_2^2}\right\} \bigg) 
\end{split}
\end{align}
where $R_t$ in \eqref{eqn:RIR-DF} shows the maximum rate at which the relay can reliably decode, while $R_f$ is the maximum rate at which relay can reliably transmit to the receiver.

\textbf{LOS scenario:} For the LOS case, the channels follow \eqref{eq:channelmodel} with $L=1$ and $\alpha_1=1$ leading to $\bh_t=\sqrt{\rho_t} \ba\left(\theta^{\mathrm{az}}_{\ell},\theta^\mathrm{el}_{\ell}\right)$. Moreover, we can expand $\xi^\circ_t = \rho_t \eta_t$ with the definition $\eta_t = \left(\frac{1}{M}\sum |[\bg_{t}]_m|\right)^2$. The spectral efficiency becomes
\begin{equation} \label{eq:dfrirupper}
\begin{split}
R_\mathrm{RIR}^\mathrm{DF} = \log_2\bigg(1+ \kappa M^2 \cdot \min\left\{\frac{p_t \rho_t \eta_t} {\sigma_1^2}, \frac{p_r \rho_r \eta_r} {\sigma_2^2}\right\} \bigg) .
\end{split}
\end{equation}

In addition, the near-field gain can be bounded by $\frac{\eta_t}{M} \leq 1$, due to the conservation of energy, resulting in 
\begin{equation}
\begin{split}
R_\mathrm{RIR}^\mathrm{DF} \leq \log_2\bigg(1+ \kappa M\cdot \min\left\{\frac{p_t \rho_t} {\sigma_1^2}, \frac{p_r \rho_r} {\sigma_2^2}\right\} \bigg).
\end{split}
\end{equation}
We note that this expression clearly indicates that proposed relay-aided IRS model can offer $\kappa M$ gain on SNR of DF-relay with a LOS path as can be seen by setting $\zeta=\rho$ in \eqref{eqn:df-relay}.
\subsubsection{AF Relay Operation}
In a similar way to \eqref{eqn:af-relay}, for the AF-relay-assisted IRS supported network, the spectral efficiency can be formulated by
\begin{align}
R_\mathrm{RIR}^\mathrm{AF} &= \log_2 \left(1+{\frac{p_t \beta \kappa^2 M^4 \xi^\circ_{t,r}}{\beta \kappa M^2 \xi_r^\circ \sigma_1^2 + \sigma^2_2}} \right)
\end{align}
for a given gain constraint 
\begin{equation} \label{eq:gainconst}
\beta \leq \frac{p_r}{p_t \kappa M^2 \xi^\circ_t + \sigma_1^2}.
\end{equation}
Moreover, with the equality of \eqref{eq:gainconst}, similarly to \eqref{eqn:af-relay2}, the expression can be simplified to
\begin{align} \label{eq:afrelaymax}
\bar{R}_\mathrm{RIR}^\mathrm{AF} &= \log_2 \left(1+\frac{\frac{\kappa M^2 p_t  \xi^\circ_t}{\sigma_1^2} \cdot \frac{\kappa M^2 p_r \xi^\circ_r}{\sigma_2^2}}{\frac{\kappa M^2 p_t \xi^\circ_t}{\sigma_1^2}+\frac{\kappa M^2 p_r \xi^\circ_r}{\sigma_2^2}+1} \right).
\end{align}
\textbf{LOS scenario:} The same channel simplifications following LOS scenario of DF-Relay allow us to form
\begin{equation}
R_\mathrm{RIR}^\mathrm{AF} = \log_2 \left(1+{\frac{p_t \beta \kappa^2 M^4 \rho_t \rho_r \eta_{t,r}}{\beta \kappa M^2 \rho_r \eta_r \sigma_1^2 + \sigma^2_2}} \right)
\end{equation}
with  $\eta_{t,r} = \left(\frac{1}{M}\sum |[\bg_{t}]_m||[\bg_{r}]_m|\right)^2$ and \eqref{eq:gainconst}. Also, maximum near-field gain $\frac{\eta_t}{M} \leq 1$ can bound the spectral efficiency as
\begin{equation}
R_\mathrm{RIR}^\mathrm{AF} \leq \log_2 \left(1+{\frac{p_t \beta \kappa^2 M^2 \rho_t \rho_r}{ \beta \kappa M \rho_r \sigma_1^2 + \sigma^2_2}} \right)
\end{equation}
since $\log(x)$, and $\frac{ax^2}{bx+c}$ for $a,b,c,x\geq0$ are strictly increasing functions. In the case of equality of gain constraint, similar expressions to \eqref{eq:afrelaymax} for only LOS path can easily be obtained.

\section{How Many Antennas Are Needed?} \label{sec:numofant}

In addition to the achievable rates, we are also interested in the number of antennas needed for providing a given gain $R_\mathrm{lim}$ over a fixed distance. To this end, we provide the expressions for the IRS, AF and DF relay-aided IRS derived from the corresponding spectral efficiency. Let us define $\gamma_\mathrm{lim}=2^{R_\mathrm{lim}}-1$ for ease of notation.

\subsection{intelligent reconfigurable Surfaces}
For the sake of a fair comparison, we consider IRS with $2M$ antennas. Therefore, the inverse function of \eqref{eqn:irsopt} for $2M$ antennas with respect to $M$ can easily be obtained as follows:
\begin{equation}
M_\mathrm{IRS} \geq \frac{1}{2} \sqrt{\frac{\gamma_\mathrm{lim} \sigma_2^2}{p_t \kappa \xi_{t,r}}} \geq \frac{1}{2} \sqrt{\frac{\gamma_\mathrm{lim}\sigma_2^2}{p_t \kappa \zeta_{t} \zeta_{r}}}
\end{equation}
Note that this is a lower bound on $M$ for an IRS with $2M$ antennas providing the rate $R_\mathrm{lim}$.

\subsection{Relay-Aided IRS}
\subsubsection{DF Relay Operation}
With relay-aided IRS with DF, the number of antennas needed to provide the gain $R_{lim}$ can be derived as
\begin{equation}
M^{DF}_\mathrm{RIR} \geq \frac{\gamma_\mathrm{lim}}{\kappa} \max \left\{ \frac{\sigma_1^2}{p_t \xi^\circ_t} , \frac{\sigma_2^2}{p_r \xi^\circ_r} \right\}
\end{equation}
using \eqref{eqn:RIR-DF}.

\subsubsection{AF Relay Operation}
The number of antennas needed for relay-aided IRS with AF depends on the gain and power limitation of the relay. Recall that we have the gain constraint \eqref{eq:gainconst} which depends on $M$. For a given amplifier coefficient $\beta$, we first find the positive solution $\tilde{M}_\mathrm{RIR}^\mathrm{AF}$ to the quadratic equation of $M^2$ given by
\begin{equation}
p_t \beta \kappa^2 \xi^\circ_{t,r} M^4 - \gamma_\mathrm{lim} \beta \kappa \xi_r^\circ \sigma_1^2 M^2 -  \gamma_\mathrm{lim} \sigma_2^2.
\end{equation}
If corresponding $\tilde{M}_\mathrm{RIR}^\mathrm{AF}$ holds for \eqref{eq:gainconst}, then the maximum gain does not violate the power constraint and ${M}_\mathrm{RIR}^\mathrm{AF}=\tilde{M}_\mathrm{RIR}^\mathrm{AF}$. Otherwise, the system applies maximum power instead of the maximum relay gain through \eqref{eq:afrelaymax} and number of antennas needed in this case can be formulated as the positive solution of the following quadratic equation of $M^2$:
\begin{equation}
\frac{\kappa p_t  \xi^\circ_t}{\sigma_1^2} \cdot \frac{\kappa p_r \xi^\circ_r}{\sigma_2^2} M^4 - \gamma_\mathrm{lim} \left(\frac{\kappa p_t \xi^\circ_t}{\sigma_1^2}+\frac{\kappa p_r \xi^\circ_r}{\sigma_2^2} \right) M^2 - \gamma_\mathrm{lim}.
\end{equation}

\section{Simulation Results}  \label{sec:Results}
In this section, we evaluate the performance of our proposed relay-aided IRS architecture using numerical simulations. 

\subsection{Simulation Setup}
\begin{figure}[t]
	\centering
	\includegraphics[width=1\columnwidth]{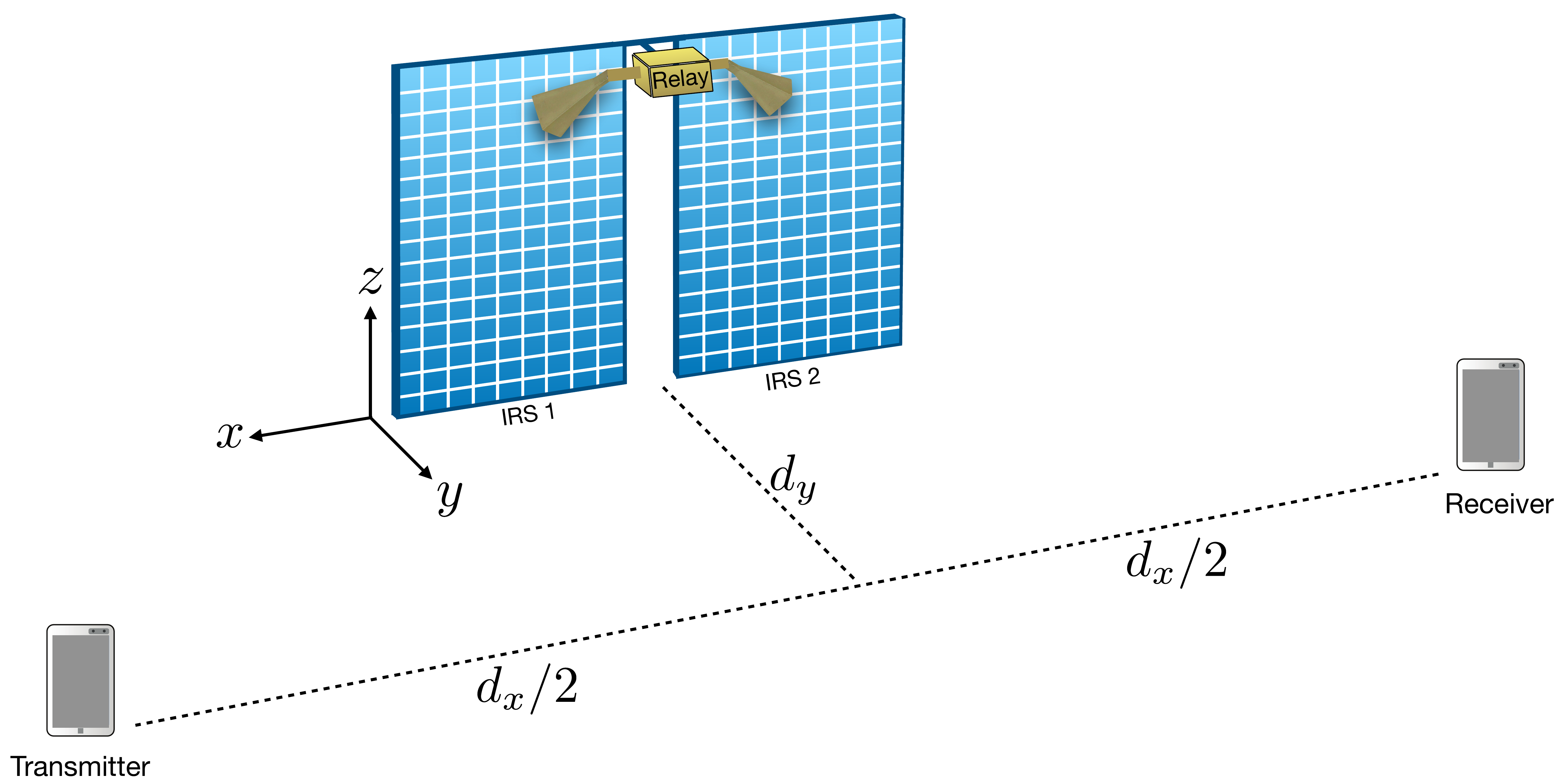}
	\caption{The adopted simulation setup where the proposed relay-aided IRS is assisting the communication between single-antenna transmitter/receiver. The IRS architecture is placed as shown in the figure with equal distances to the transmitter and receiver.}
	\label{fig:systemgeometry}
\end{figure}

We consider the scenario illustrated in \figref{fig:systemgeometry} where transmitter and receiver are located at two points aligned on $y$-axis with a separation $d_x$ on $x$-axis. The IRS/Relay/Relay-aided IRS is placed at $d_y=10m$ away from the transmitter and receiver in $y$-axis while it is in the middle of them in $x$-axis. We take the heights of transmitter and IRS/Relay/Relay-aided IRS units as $10m$ and the receiver as $1m$. In this setup, the channel gains are generated by using 3GPP Urban Micro (UMi) - street canyon model \cite{3GPP2017} given as 
$$P_\mathrm{loss} = 32.4+21\log_\mathrm{10}(d_\mathrm{3D})+20\log_\mathrm{10}(f_c)$$
where $d_\mathrm{3D}$ and $f_c$ denote the 3D LOS path distance in meters and carrier frequency in GHz, respectively. In the following simulations, we consider the LOS scenario with the near-field upper-bounds. The LOS channel gains $\rho_r$ and $\rho_t$ are computed with UMi model and utilized in the achievable rates of IRS, DF and AF relays, and the upper-bounds for relay-aided IRS with DF and AF through the equations derived in Section \ref{sec:rates} and \ref{sec:numofant}.

As detailed earlier, the IRS considers twice the size of reflection elements $2M$ compared to as there are two IRSs adopted in relay-aided IRS. We consider a transmitter power of $p_t=20$dBm and a relay maximum power of $p_r=20$dBm for all scenarios.  We consider two different carrier frequency values $60$ GHz and $3.5$ GHz representing mmWave and sub-6 GHz channels. The noise figure is set at $8$dB and the bandwidth is assumed to be $100$ MHz at the 3.5GHz band and $1$GHz at the $60$GHz band. A unitary reflection coefficient, $\alpha=1$, is adopted assuming perfect reflection at all the relay-aided IRS/IRS surfaces. For the simulations where AF relay gain is given by $\beta$, the relays apply the minimum of $\beta$ or the amplification gain using maximum power.

\subsection{Achievable Rates}
\begin{figure}[t]
	\centering
	\includegraphics[width=1\columnwidth, height=170pt]{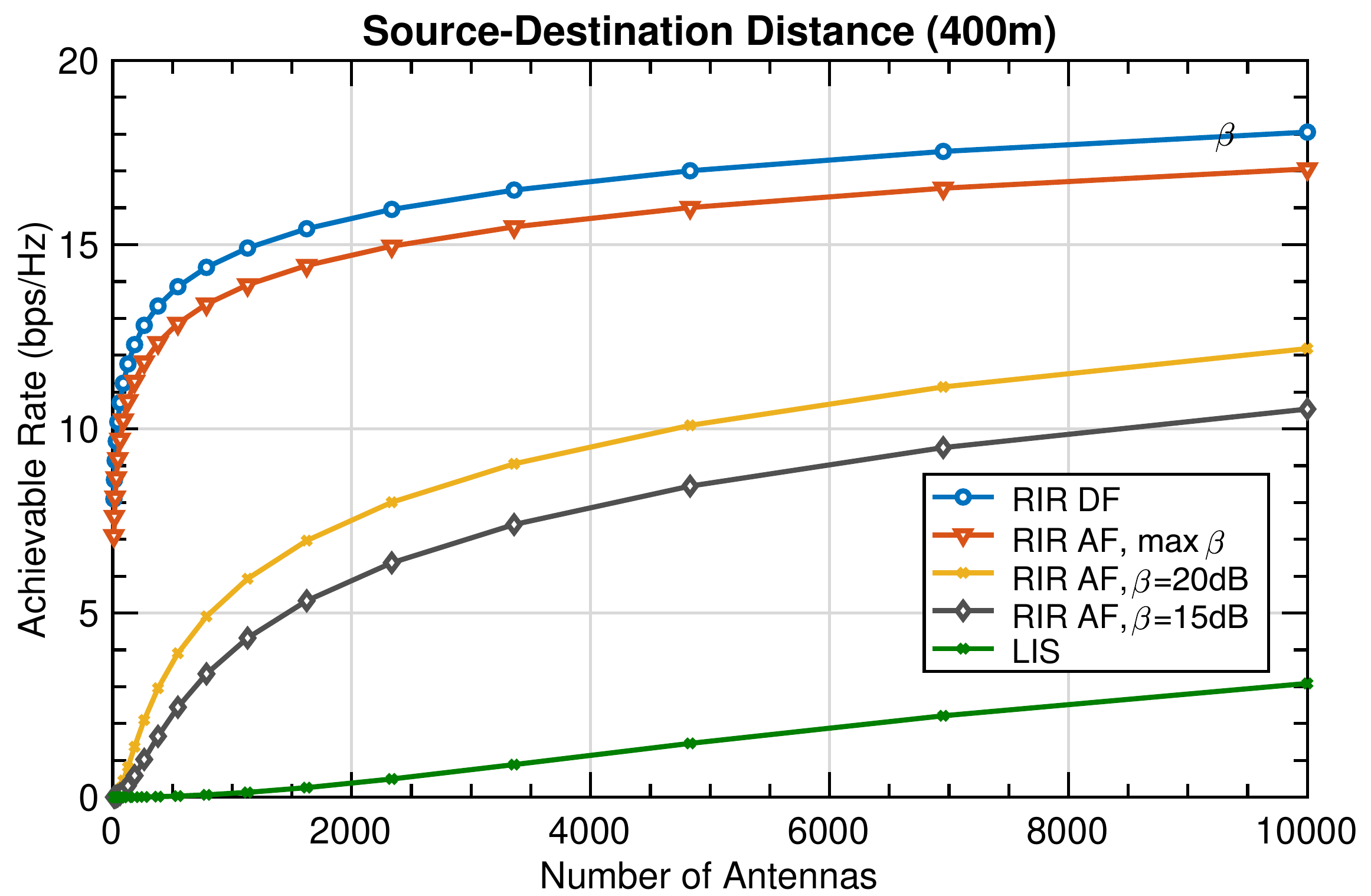}
	\caption{The achievable rates of the proposed relay-aided IRS architectures versus the classical intelligent reconfigurable surfaces (IRS) at different numbers of surface antennas. $M$. The setup operates at $3.5$GHz and assumes a fixed distance $400$m between the transmitter and receiver. }
	\label{fig:rate3p5}
\end{figure}

We first investigate the achievable rates for varying number of antennas $M$ and over a fixed distance $d_x=400m$ between the transmitter and receiver. In \figref{fig:rate3p5}, the achievable rates with respect to the number of antennas are plotted considering a setup operating at  $3.5$ GHz and with a bandwidth $100$MHz. As shown in this figure, the proposed relay-aided IRS architectures achieve much higher spectral efficiency compared to the classical intelligent reconfigurable  at any fixed number of antennas. Further, this figure illustrates that relay-aided IRS with a DF achieves higher gain compared to the relay-aided IRS with AF relay (for the case of maximum used $\beta$). DF relays, however, require relatively higher hardware complexity and latency (initial offset) overhead which is the cost of the higher achievable rate.

\begin{figure}[t]
	\centering
	\includegraphics[width=1\columnwidth, height=170pt]{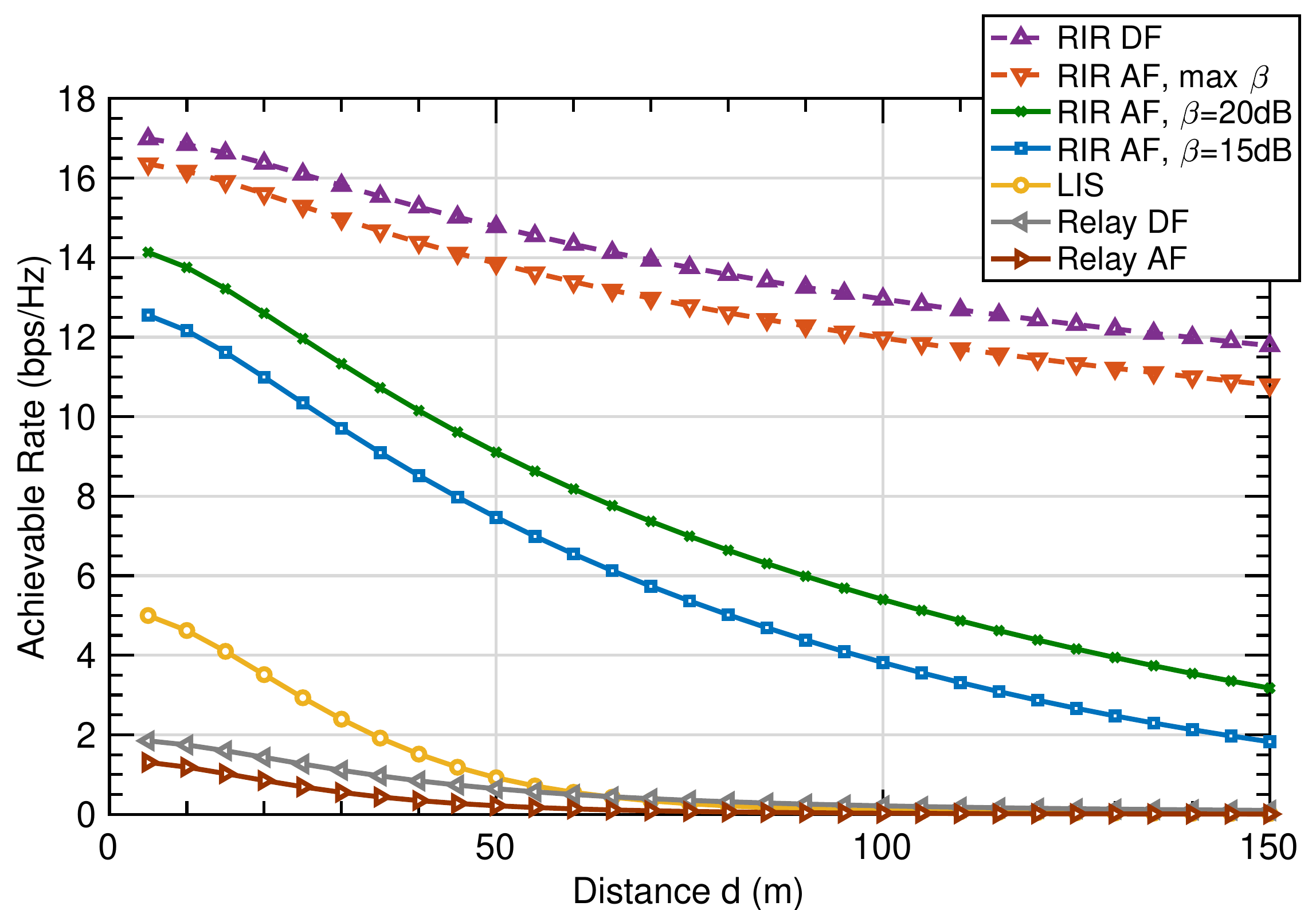}
	\caption{The achievable rates of the proposed relay-aided IRS architectures versus the classical intelligent reconfigurable surfaces (IRS) and the standard single-antenna AF/DF relays at different distances between the transmitter and receiver. The setup operates at $60$GHz and assumes a fixed number of elements at the surfaces, $M=50$ thousands. }
	\label{fig:rate60}
\end{figure}

\figref{fig:rate3p5} also plots the achievable rates with the proposed relay-aided IRS architecture with AF relays under different realistic values for the amplification gains $\beta$. In general, however, the relay-aided IRS with AF and reasonable amplification gain results in better performance compared to the classical IRS. This is because the IRS requires a massive amount of antennas to provide acceptable SNR gains, while the proposed relay-aided IRS architecture splits the target SNR gain between the number of elements and the amplification gain. At the $60$GHz band, the achievable rates using these different architectures are  evaluated in \figref{fig:rate60} for different values of the distance between the transmitter and receiver. This figure emphasizes the potential gain of the proposed architectures compared to both classical intelligent reconfigurable surfaces and standard single-antenna relays.

\subsection{How Many Elements Do We Need?}
\begin{figure}[t]
	\centering
	\includegraphics[width=1\columnwidth, height=170pt]{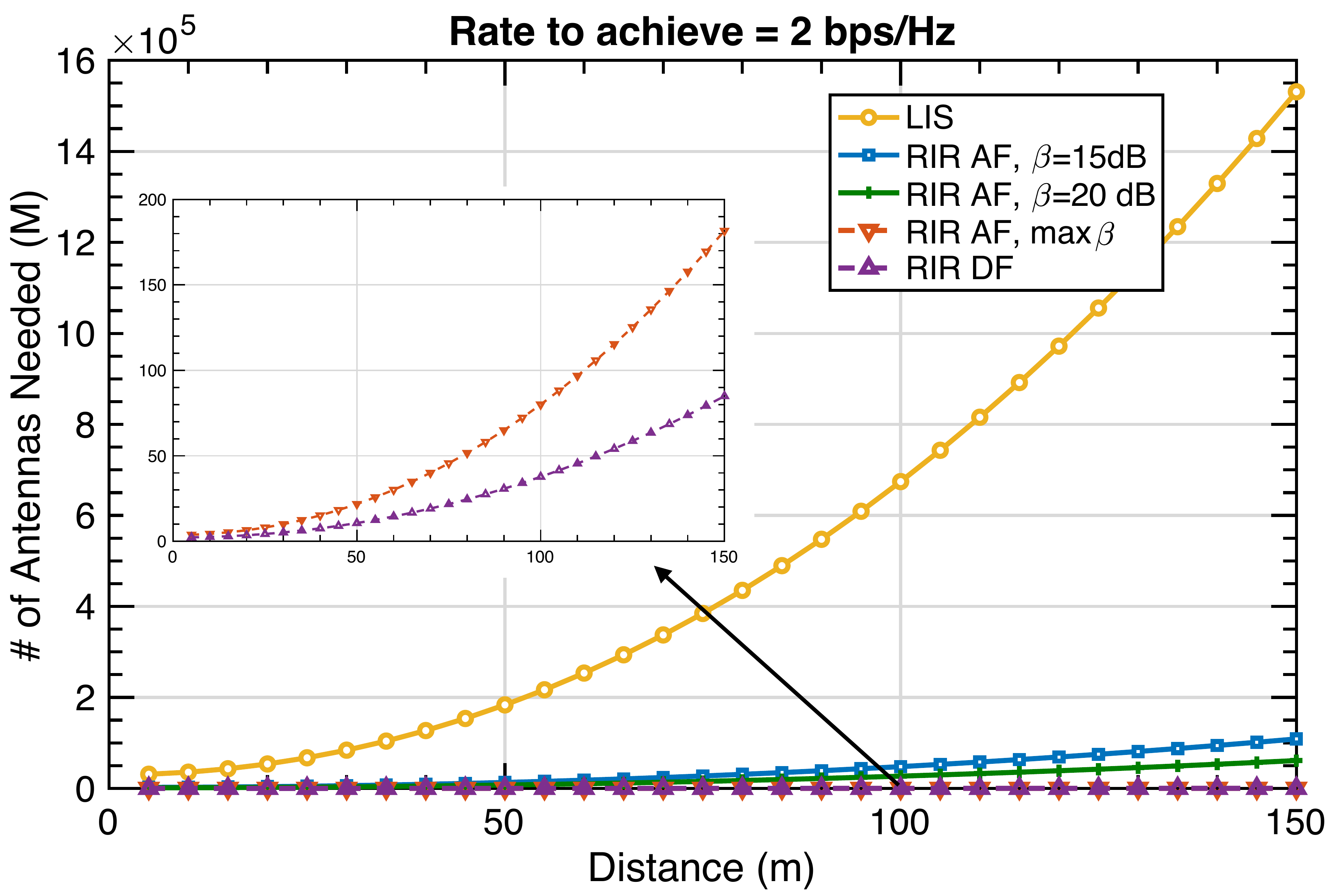}
	\caption{This figure shows the required number of antenna elements, $M$, by the proposed relay-aided IRS architectures and the classical IRS to achieve a target spectral efficiency of $2$bps/Hz. The antenna numbers are evaluated at different distances between the transmitter and receiver which operate at $60$GHz frequency band. }
	\label{fig:ant60}
\end{figure}

We next examine the number of elements needed to provide a fixed rate for varying distances between the transmitter and receiver. Note that distance between the transmitter-IRS and IRS-receiver also increases with the increasing distance as shown in \figref{fig:systemgeometry}. \figref{fig:ant60} shows the required number of antenna elements  providing a fixed target rate $R_\mathrm{lim}=2$ bps/Hz at $60$ GHz carrier frequency. The number of elements needed scales exponentially for IRS and much larger than all relay-aided IRS architectures.With the classical IRS architecture, the required number of  elements easily exceeds $100$ thousands over $25$m. On the other hand, \figref{fig:ant60}, shows that this number may be required by the relay-aided IRS with AF architecture at a distance 150m and amplification gain $\beta=15$dB. Increasing the amplification gain to $20$dB can further reduce this number to nearly $50$ thousand elements. Further, this figure shows that the proposed relay-aided IRS architecture with DF relay may need much less number of elements. At $150$m, only $100$ elements per surface are needed for the relay-aided IRS with DF to achieve the same target SNR.  In general, \figref{fig:ant60} shows that the proposed relay-aided IRS architectures can significantly reduce the required number of elements to achieve a reasonable achievable rate target at different distances yielding a promising solution for practical deployments of intelligent reconfigurable surfaces.

\section{Conclusion}
In this paper, we proposed a novel design for the intelligent reconfigurable surfaces based on connecting two intelligent surfaces via a full-duplex relay. By combining the standard single-antenna relay and the intelligent reconfigurable surfaces, the new relay-aided IRS design has adopted the beamforming function of IRS devices while maintaining the active terminal nature of the traditional relay. We developed an accurate channel model by combining both the far-field channels between the transmitter/receiver and the IRS and the near-field channels between the intelligent surfaces and the relay. Further, we derived closed-form expressions for the achievable rates using the proposed relay-aided IRS architecture and for how many antennas are needed compared to classical IRS approaches.  Numerical simulations showed that the proposed relay-aided IRS architecture outperforms both the IRS and the standard relay solutions with much higher achievable rate, particularly at high frequencies. Compared to traditional IRS solutions, the relay-aided IRS architecture can achieve the same transmission rate with much fewer elements (and hence with less training overhead and more robustness). Further, since the proposed architecture has both the amplification and beamforming capabilities, it has clear advantages over MIMO relays, which could be as costly as the base stations. As the relay-aided IRS architecture are consisted of two reflect arrays connected via an optical fiber, it can be deployed in very flexible ways: The relay-aided IRS reflect arrays can be installed separately, on different faces of the building or walls, or can be tilted to look at different directions. This has the potential of enhancing the coverage gains provided by the proposed architectures.  All these interesting gains yield the proposed relay-aided IRS architecture as a promising solution for practical intelligent surface deployments.



\balance

\end{document}